# ПЕРИОДИЧЕСКИЕ МЕРЫ ГИББСА ДЛЯ МОДЕЛИ ПОТТСА-SOS НА ДЕРЕВЕ КЭЛИ.

## Рахматуллаев М.М., Расулова М.А.

Пусть $\tau^k = (V, L)$, $k \geq 1$ есть дерево Кэли порядка $k$, где $V$ - множество вершин, $L$ - множество ребер $\tau^k$. Известно, что $\tau^k$ можно представить как $G_k$ – свободное произведение $k+1$ циклических групп второго порядка (см. [1]).

Мы рассматриваем модели, где спин принимает значения из множества $\Phi = \{1, 2, ..., m\}$, $m \geq 1$. Тогда конфигурация $\sigma$ на $V$ определяется как функция $x \in V \to \sigma(x) \in \Phi$; множество всех конфигураций совпадает с $\Omega = \Phi^V$.

Гамильтониан модели Поттса-SOS определяется как следующем образом (см. [5]):

$$H(\sigma) = -J \sum_{<x,y>\in L} |\sigma(x) - \sigma(y)| - J_p \sum_{<x,y>\in L} \delta_{\sigma(x)\sigma(y)}, \qquad (1)$$

где $J, J_p \in R$ $<x, y>$ – ближайшие соседи, $\delta_{uv}$ – символ Кронекера.

В случае $J_p = 0$, $J \neq 0$ модель (1) совпадает с моделью SOS. Трансляционно-инвариантные и периодические меры Гиббса для модели SOS были изучены в работах [1], [2]. В случае $J = 0$, $J_p \neq 0$ модель (1) совпадает с моделью Поттса. Для модели Поттса трансляционно-инвариантные меры Гиббса описаны в работе [3], а в работах [4], [6] были изучены периодические меры Гиббса.

В работе [5] изучены трансляционно-инвариантные меры Гиббса для модели Поттса-SOS. В настоящей работе изучаются периодические меры Гиббса для гамильтониана (1) на дереве Кэли порядка два.

Известно [5], что каждой мере Гиббса для модели Поттса-SOS на дереве Кэли порядка $k \geq 1$, можно сопоставить совокупность векторов

$$h_x^* = \left( h_{0,x}, \ h_{1,x}, ..., \ h_{m-1,x} \right), \quad x \in G_k,$$

удовлетворяющих уравнениям

$$h_x^* = \sum_{y \in S(x)} F\left(h_y^*, m, \theta, r\right), \tag{2}$$

где $S(x)$ - множество «прямых потомков», точки $x \in G_k$ и $\theta = \exp(J\beta)$, $r = \exp(J_p \beta)$, $\beta = \dfrac{1}{T}$, $T > 0$, и функция $F(\cdot, m, \theta, r) : R^m \to R^m$ определена следующим образом: $F(h, m, \theta, r) = \left( F_0(h, m, \theta, r), ..., F_{m-1}(h, m, \theta, r) \right)$ с

$$F_i\left(h, m, \theta, r\right) = \ln \left( \dfrac{\sum_{j=0}^{m-1} \theta^{|i-j|} r^{\delta_{ij}} \exp(h_j) + \theta^{m-i} r^{\delta_{mi}}}{\sum_{j=0}^{m-1} \theta^{m-j} r^{\delta_{mj}} \exp(h_j) + r} \right), \tag{3}$$

где $h = (h_0, h_1, ..., h_{m-1})$, $i = 0, 1, 2, ..., m-1$.

Пусть $G_k / \overline{G}_k = \{H_1, H_2, ..., H_s\}$ – фактор-группа, где $\overline{G}_k$ – нормальный делитель индекса $s \geq 1$.

**Определение 1.** Совокупность векторов $h = \{h_x, x \in G_k\}$ называется $\overline{G}_k$-периодической, если $h_{xy} = h_x$ для любого $x \in G_k$, $y \in \overline{G}_k$.

$G_k$-периодическая совокупность векторов называется трансляционно-инвариантной.

**Определение 2.** Мера $\mu$ называется $\overline{G}_k$-периодической (трансляционно-инвариантной), если она соответствует $\overline{G}_k$-периодической (трансляционно-инвариантной) совокупности векторов $h$.

Пусть $m = 2$. Рассмотрим нормальный делитель индекса два $G_k^{(2)} = \{x \in G_k : |x| - \text{четно}\}$ и фактор-группу $G_k / G_k^{(2)} = \{G_k^{(2)}, \ G_k \setminus G_k^{(2)}\}$. Тогда $G_k^{(2)}$ – периодическая совокупность векторов имеет следующий вид:

$$\tilde{h}_x = \begin{cases} h_x, & \text{если} \quad x \in G_k^{(2)}, \\ l_x, & \text{если} \quad x \in G_k \setminus G_k^{(2)}, \end{cases}$$

где $h_x = (h_{0,x}, h_{1,x})$, $l_x = (l_{0,x}, l_{1,x})$.

Введем обозначения $z_0 = \exp(h_{0,x})$, $z_1 = \exp(h_{1,x})$, $t_0 = \exp(l_{0,x})$, $t_1 = \exp(l_{1,x})$, тогда из (2) и (3) имеем следующую систему уравнений:

$$\begin{cases} z_0 = \left( \dfrac{rt_0 + \theta t_1 + \theta^2}{\theta^2 t_0 + \theta t_1 + r} \right)^k, \\ z_1 = \left( \dfrac{\theta t_0 + rt_1 + \theta}{\theta^2 t_0 + \theta t_1 + r} \right)^k, \\ t_0 = \left( \dfrac{rz_0 + \theta z_1 + \theta^2}{\theta^2 z_0 + \theta z_1 + r} \right)^k, \\ t_1 = \left( \dfrac{\theta z_0 + rz_1 + \theta}{\theta^2 z_0 + \theta z_1 + r} \right)^k, \end{cases} \quad (4)$$

где $\theta = \exp(J\beta)$, $r = \exp(J_p\beta)$.

Из первого и третьего уравнений системы (4) имеем

$$\begin{cases} \sqrt[k]{z_0} - 1 = \dfrac{(z_0 - 1)(r - \theta^2)}{\theta^2 t_0 + \theta t_1 + r}, \\ \sqrt[k]{t_0} - 1 = \dfrac{(t_0 - 1)(r - \theta^2)}{\theta^2 z_0 + \theta z_1 + r}. \end{cases} \quad (5)$$

Отсюда получим, что $(z_0, t_0) = (1, 1)$ является решением системы уравнений (5) для любых $\theta, r, z_1, t_1$. В этом случае из второго и четвертого уравнений системы (4) имеем следующее:

$$\begin{cases} z_1 = \left( \dfrac{2\theta + rt_1}{\theta^2 + \theta t_1 + r} \right)^k, \\ t_1 = \left( \dfrac{2\theta + rz_1}{\theta^2 + \theta z_1 + r} \right)^k. \end{cases} \quad (6)$$

Введем обозначение $f(z_1) = \left( \dfrac{2\theta + rz_1}{\theta^2 + \theta z_1 + r} \right)^k$.

Тогда система уравнений (6) сводится к следующему уравнению:
$$f(f(z_1)) - z_1 = 0. \qquad (7)$$

Ясно, что решение уравнения $f(z_1) = z_1$ является решением для уравнения (7). Нас интересуют решения уравнения (7), отличные от решения уравнения $f(z_1) = z_1$. Эти решения соответствуют $G_k^{(2)}$–периодическим мерам Гиббса, которые не являются трансляционно-инвариантными. Отбрасывая решения уравнения $f(z_1) = z_1$ из решений уравнения (7) в условии $k = 2$, и упростив, получаем следующее квадратное уравнение:

$$\left(\theta^6 + 2\theta^4 r + \theta^2 r^2 + r^4 + 2\theta r^3 + 2\theta^3 r^2\right) z_1^2 +$$
$$+ \left(2\theta^7 + 6\theta^5 r + 6\theta^3 r^2 + 6\theta r^3 - 4\theta^4 + \theta^4 r^2 + 8\theta^2 r^2 + 2\theta^2 r^3 + 8\theta^4 r + r^4\right) z_1 +$$
$$+ 4\theta^2 r^2 + 4\theta^6 r + r^4 + 6\theta^4 r^2 + 4\theta^2 r^3 + \theta^8 + 4\theta^5 r + 8\theta^3 r^2 + 4\theta r^3 = 0.$$

Чтобы это уравнение имело два действительных положительных корня, требуется выполнение условий $D > 0, b < 0$, где

$$D = \left(2\theta^7 + 6\theta^5 r + 6\theta^3 r^2 + 6\theta r^3 - 4\theta^4 + \theta^4 r^2 + 8\theta^2 r^2 + 2\theta^2 r^3 + 8\theta^4 r + r^4\right)^2 -$$
$$\left(\theta^6 + 2\theta^4 r + \theta^2 r^2 + r^4 + 2\theta r^3 + 2\theta^3 r^2\right) \cdot (4\theta^2 r^2 + 4\theta^6 r + r^4 + 6\theta^4 r^2 + 4\theta^2 r^3 +$$
$$\theta^8 + 4\theta^5 r + 8\theta^3 r^2 + 4\theta r^3),$$
$$b = 2\theta^7 + 6\theta^5 r + 6\theta^3 r^2 + 6\theta r^3 - 4\theta^4 + \theta^4 r^2 + 8\theta^2 r^2 + 2\theta^2 r^3 + 8\theta^4 r + r^4.$$

**Теорема**. Пусть $k = 2$. Если $D > 0, b < 0$, тогда существуют не менее двух $G_k^{(2)}$–периодических (не трансляционно-инвариантных) мер Гиббса для модели Поттса-SOS; Если $D = 0, b < 0$, тогда существуют не менее одной $G_k^{(2)}$–периодической (не трансляционно-инвариантной) меры Гиббса для модели Поттса-SOS.

Отметим, что множество $\{(r, \theta) \in R^2 : D \geq 0, b < 0\}$ является непустым. Действительно, пусть $r = \theta^2$. Тогда имеем

$$D = -16\theta^8 \left(\theta^2 - 1\right)^2 \left(3\theta^4 + 10\theta^3 + 6\theta^2 - 1\right),$$
$$b = 4\theta^4 \left(\theta^4 + 5\theta^3 + 4\theta^2 - 1\right).$$

Если $\theta < \theta_D$ $(\theta_D \approx 0{,}32359)$, тогда имеет место $D > 0, b < 0$, т.е. существуют не менее двух $G_k^{(2)}$–периодических (не трансляционно-инвариантных) мер Гиббса; Если $\theta=\theta_D$, тогда выполняется следующее соотношение $D=0, b<0$, т.е. существуют не менее одной $G_k^{(2)}$ – периодической (не трансляционно-инвариантной) меры Гиббса.

**Замечание.** Заметим, что при $k=2$ для модели Поттса не существуют периодические меры Гиббса, но для модели Поттса-SOS при некоторых условиях такая мера существует.

## ЛИТЕРАТУРЫ